\documentclass[aip,reprint]{revtex4-2}
\usepackage{graphicx}
\usepackage{color}

\begin{document}

\title{Thermo-kinetic explosions: \emph{safety first} or  \emph{safety last}? 
 }

\author{Julyan H. E. Cartwright}
\email{julyan.cartwright@csic.es}
\affiliation{Instituto Andaluz de Ciencias de la Tierra, CSIC--Universidad de Granada, 18100 Armilla, Granada, Spain}
\affiliation{Instituto Carlos I de F\'{\i}sica Te\'orica y Computacional,  Universidad de Granada,  18071 Granada, Spain}

\date{Version of \today}

\begin{abstract}
Gas and vapour explosions have been involved in industrial accidents since the beginnings of industry. A century ago, 
at 11:55 am on Friday 24th September 1920, the petroleum barge \emph{Warwick} exploded in London's docklands and seven men were killed.
Understanding what happened when it blew up as it was being refurbished, and how to prevent similar explosions, involves  fluid mechanics and thermodynamics plus chemistry.  I recount the 1920 accident as an example, together with the history of thermo-kinetic explosions prior to 1920 and up to the present day, and I review the history and the actual state of the science of explosion and  
the roles of fluid mechanics, thermodynamics, and chemistry in that science.
The science of explosions has been aware of its societal implications from the beginning, but, 
despite advances in health and safety over the past century, is there still work to do?
\end{abstract}

\maketitle

\section{Gas and vapour explosions before 1920}

\noindent
One risk of using a public toilet in ancient Rome was of getting one's bottom singed (or worse);  methane explosions in sewers were known from classical times \cite{scobie1986,koloski2015,deming2020}. Mines too have always accumulated inflammable gases, in particular firedamp --- methane ---, which has caused explosions for as long as mining itself. So it is not surprising that it is in mining that we see the first research on preventing gas explosions. 

The 1812 disaster at Felling pit near Gateshead, where 92 men were killed through a methane explosion, was one of the stimuli for the invention of safety lamps to prevent explosions \cite{morgan1936}.
Humphry Davy and George Stephenson were two of the key figures involved in developing safe miners' lamps --- the Davy lamp  \cite{davy1816} and the Geordie lamp \cite{brandling1817} respectively ---  designed to supersede the use of naked candle flames to illuminate mines. 
Davy's lamp used a wire gauze to supply air; Stephenson's Geordie lamp used narrow tubes for the same purpose. There was a rather squalid dispute between the two men and their supporters over priority in the invention \cite{morgan1936,brandling1817}.

Humphry Davy had an assistant when working on the safety lamp. That was Michael Faraday. Faraday later took over Davy's position at the Royal Institution and wrote a beautiful book, \emph{The Chemical History of a Candle} \cite{faraday}. About  mine explosions, he wrote that 
\begin{quote}
``In olden times the miner had to find his own candles, and it was supposed that a small candle would not so soon set fire to the fire-damp in the coal mines as a large one ... They have been replaced since then by the steel-mill, and then by the Davy-lamp, and other safety-lamps of various kinds.''
\end{quote}
Early safety lamps were dimmer than candles and the gauze or tube did not completely eliminate the risk of explosion, and improvements to the designs of miners' safety lamps continued to be worked on throughout the 19th century \cite{morgan1936,thomas2015}.

As well as gases, inflammable vapours\footnote{There is no difference between a gas and a vapour in terms of their fluid physics or chemistry; only in terms of their thermodynamics. The different terms simply indicate that a gas is a substance  found at that temperature and pressure only in the single thermodynamic gaseous state, whereas a vapour is also found in its condensed phase at the same temperature and pressure.} were likewise the cause of explosions.
Thomas Graham, of Graham's law fame, was one of those who investigated the loss of the paddle steamer \emph{Amazon} in 1852. While the cause of the fire that led to 105 to 115 deaths when she sank off the Isles of Scilly on her maiden voyage was never absolutely established, Graham thought that the presence of volatile inflammable liquids in the store room near to the engines was a key factor \cite{graham1853}: 
\begin{quote}``The sudden and powerful burst of flame from the store-room, 
which occurred at the very outset of the conflagration, suggests 
strongly the intervention of a volatile combustible, such as 
turpentine, although the presence of a tin can of that substance in 
the store-room appears to be left uncertain. I find, upon trial, 
that the vapour given off by oil of turpentine is sufficiently 
dense, at a temperature somewhat below 100$^\circ$, to make air explosive upon the approach of a light. 
Any escape of turpentine 
from the heated store-room would therefore endanger a spread of 
flame by the vapour communicating with the lamps burning in 
the boiler room or even with the fires of the furnaces.'' 
\end{quote}

On 2nd October 1874, the \emph{Tilbury}, a barge on the Regent's Canal in London,  exploded killing its crew of four; it was carrying both barrels of petroleum and gunpowder. The spot, at the edge of Regent's Park under a bridge over the canal, is now called \emph{Blow Up Bridge}. It was lucky, as it  were,  that this happened at 5am, otherwise many more people would have died \cite{taylor1874,holland1874}.    Frederick Abel, developer of guncotton and inventor of cordite,
mentioned this example  in a lecture ``On Accidental Explosions'' he gave at the Royal Institution in 1875 that was subsequently printed in Nature \cite{abel1875}:
\begin{quote}
``Among other ``accidents'' referred to as arising from a similar cause, was the recent explosion of the powder-laden barge in the Regent's Canal. It was established by a sound chain of circumstantial evidence that this explosion must have been caused by the ignition, in the cabin of the barge, of an explosive mixture of air and of the vapour of petroleum, derived from the leakage of certain packages of the spirit which were packed along with the powder.''
\end{quote}
The  \emph{Tilbury} accident accelerated the passing of the Explosives Act in 1875, which in the UK regulated the manufacture and carriage of dangerous substances.

The industrializing world was demanding more and more chemical products. The petrochemical industry was, figuratively speaking, exploding. Abel returned to the theme of accidental explosions a decade later for another lecture at the Royal Institution in 1885, this time entitled ``Accidental Explosions Produced by Non-Explosive Liquids'', in which he specifically discussed ``accidents connected with the transport, storage, and use of volatile inflammable liquids which are receiving extensive application, chiefly as solvents and as illuminating agents'' \cite{abel1885}. 

Redwood wrote a very thorough report in 1894 on `the transport of petroleum in bulk' \cite{redwood1894} motivated by the explosion of the petroleum tank-steamship \emph{Tancarville} while being worked on in Newport, Wales in 1891, in which 5 men were killed.  Redwood comments
\begin{quote}
``In a lecture delivered at the Royal Institution, on the 12th March, 1875, Sir Frederick Abel called attention to the special danger arising from the accumulation of the vapour of petroleum spirit, or or the similar liquid, in unventilated places and
referred to several cases of fire and explosion in illustration of his remarks, including one which occurred with coal-tar naphtha in 1847. In a more recent lecture (on the 13th March, 1885) he enlarged upon this theme, and gave particulars of explosions due to the use on board Her Majesty's ships of paint-driers containing dangerously inflammable liquid hydrocarbons. For many years past Colonel Majendie has briefly described in the Annual Reports of H.M.\ Inspectors of Explosives, the more
noteworthy of the petroleum accidents which have taken place during the preceding twelvemonths (although petroleum, not being an explosive substance, does not come within the scope of the Explosives Act); and his reports constitute most valuable contributions to the literature of the subject.''
\end{quote}

In his report Redwood lists many other similar incidents that give a depressing litany even at this early date of the petrochemical industry. 
\begin{quote}
``These accidents arose from the incautious handling of a material which was not generally known to contain a liquid readily converted into vapour, and in that condition capable of forming a powerfully explosive mixture with air'', 
\end{quote}
he writes.
Of the \emph{Tancarville} in particular he relates 
\begin{quote}
``The evidence given at the Board of Trade
inquiry conclusively demonstrated that the accident was due to the ignition of an explosive mixture of petroleum vapour and air in the ballast tank, but did not clear up the question of how the ignition took place. Samples of the petroleum were examined by Dr.\ Dupr\'e and by the Author, with the result that it was ascertained that one gallon of such oil would render 200 cubic feet of air feebly, and, about 58 feet of air strongly, explosive; therefore 20 cubic feet of oil would have sufficed to render the atmosphere of the water-ballast tank, the capacity of which was about 6,000 cubic feet, explosive.'' 
\end{quote}

As we shall see, the 1920 accident was to be very similar in its origin. 
Just the year before the accident that is our principal example, on 15th July 1919 there occurred the Cardiff dockland disaster when the oil tanker \emph{Roseleaf} exploded, killing the 12 men working on her. This was determined to be owing to a man carrying a naked flame down into the ship.

\section{The explosion of the petroleum barge, \emph{Warwick}, on 24th September 1920}

\begin{figure}[tbp]
\centering\includegraphics*[width=\columnwidth,clip=true]{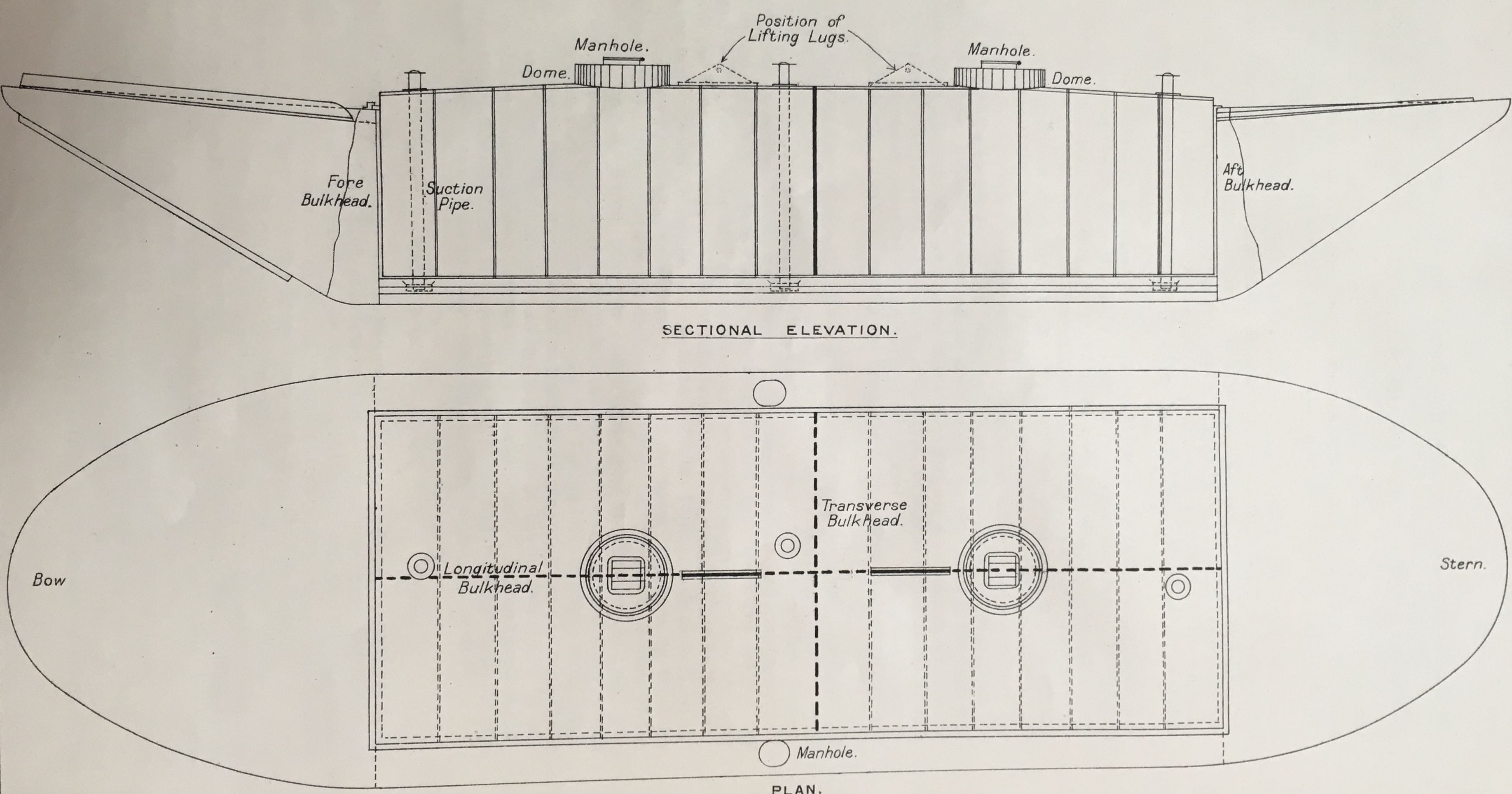} \\
\centering\includegraphics*[width=\columnwidth,clip=true]{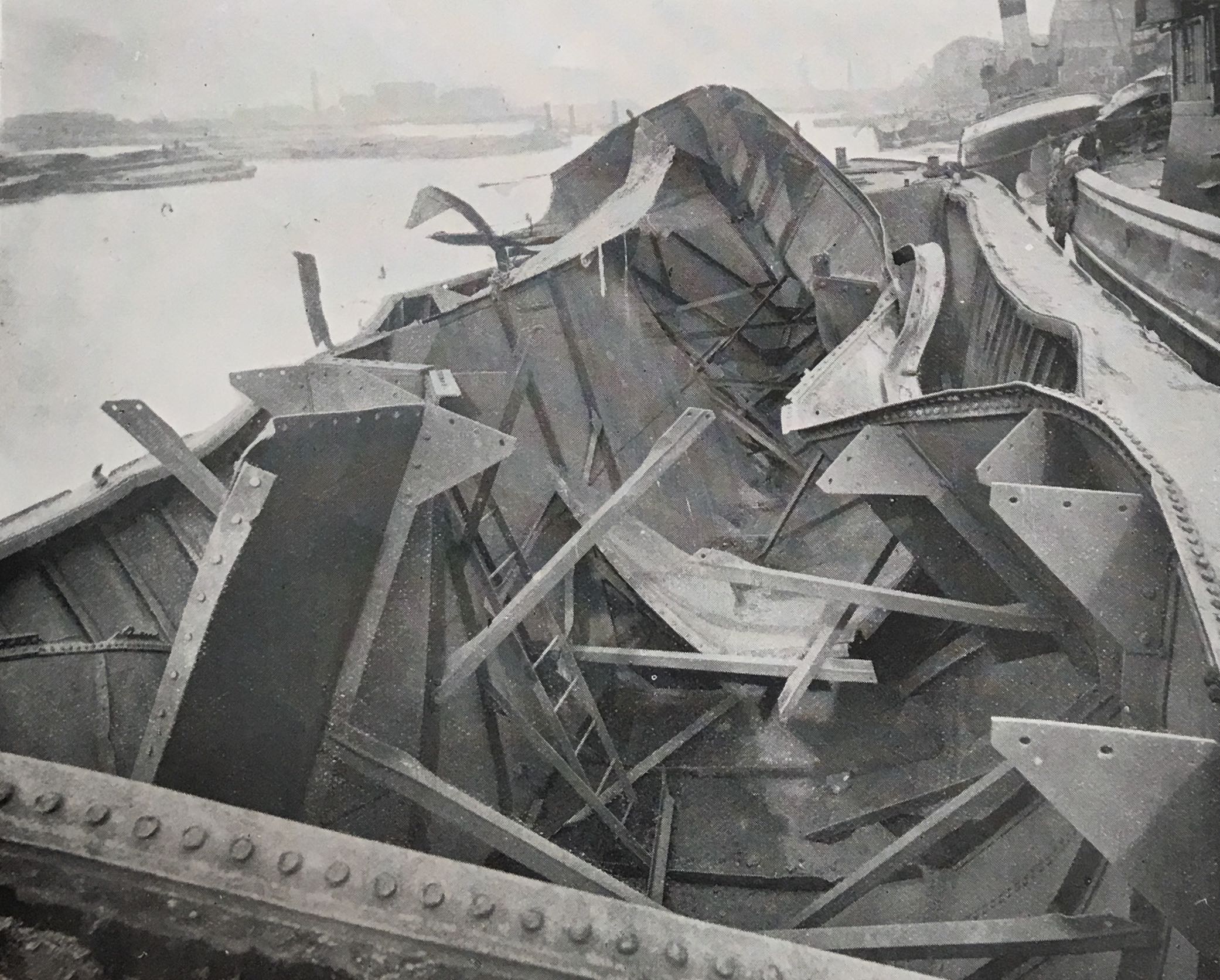} \\
\centering\includegraphics*[width=\columnwidth,clip=true]{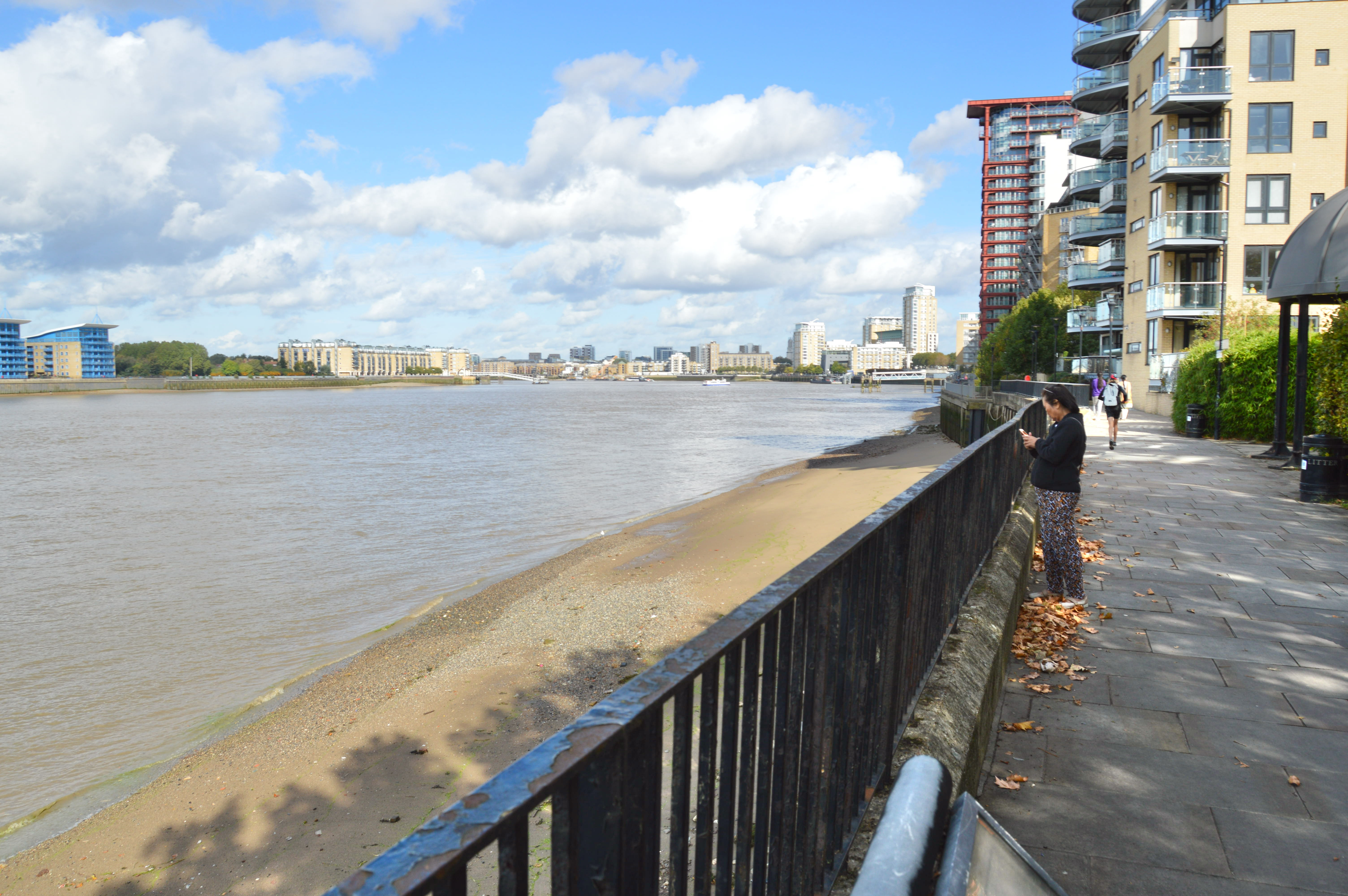}
\caption{\label{fig:warwick}
The explosion of the petroleum barge, \emph{Warwick}, on 24th September 1920.
Above: the design of the  barge, showing its petrol tank.
Centre: the remains of the \emph{Warwick} after the explosion. Both images reproduced from the accident report \cite{taylor1921}.
Below: the same view a century later, on 24th September 2020; the Thames is now a post-industrial river.
}
\end{figure}

The Isle of Dogs, a tongue of land formed by a large curve of the River Thames in east London, was once marshes but  by the turn of the 20th century had become a great hub of industrial activity related to the river. Samuel Hodge \& Sons, engineers and ship repairers, had premises  at Union Iron Works, 104 Westferry Road in Millwall, the Isle of Dogs, from the 1850s to the mid-1920s. The site of these works where the accident that killed 7 men occurred is now part of a public park on the riverbank, the Sir John McDougall Gardens.
My grandfather, Charles James Cartwright (1882--1920) travelled across the river from Bermondsey to the Isle of Dogs every day. He worked as a welder at Samuel Hodge's until his death in this accident alongside his brother, who was also killed, and his father-in-law, my great-grandfather,  Thomas Tilling, who survived the explosion.
The following account of the accident is taken from the report written by G. Stevenson Taylor, Inspector of Factories  \cite{taylor1921}.

The \emph{Warwick} was a petroleum barge, designed to carry 42\,000 gallons of petrol in its tank (Fig.~\ref{fig:warwick}a). It had last been used for that purpose from 17th to 20th September 1920, before being sent for cleaning prior to work being carried out on it. On 22nd September it was cleaned out by pumping out the remaining 19 gallons of petrol and mopping up the remainder with cotton rags. Then on 23rd September it underwent steam cleaning by putting a steam hose into the tank.  On the morning of 24th September the barge was towed up the river from Purfleet where the cleaning had been carried out to the Isle of Dogs.
 As it approached the wharf it could be seen from the barge that a work party was ready to come aboard with acetylene torches; clearly they were intending to get to work straight away, and this alarmed the lighterman (bargeman) on board. My great-grandfather
``Tilling heard Lazell, the lighterman, tell James [the foreman] that, if they were going to use lights [i.e, flames] on the barge, he was off.'' The barge was moored at 11:45.

The idea was to remove the petrol tank from the barge to alter its bulkheads according to the latest regulations. For this, the first job was to fit lifting lugs onto the tank, which entailed removing rivets from the tank top with welding apparatus. As Taylor's report states 
\begin{quote}
``Charles Cartwright (deceased) was using an oxy-acetylene burner, and with this he was seen to burn the heads from a number of rivets along the central joint in the tank top.'' 
\end{quote}
At 11:55, the barge exploded  (Fig.~\ref{fig:warwick}b); 
``all the witnesses describe the explosion as of great violence.'' 
One of these witnesses was Thomas Tilling, who 
``was on the wharf at the time of the explosion and was blown against a wall and slightly injured.''

Let us continue with Taylor, who pieces together the facts of the matter in the best detective tradition. 
\begin{quote}
``The nature and extent of the damage to the tank and barge, as well as the evidence of the witnesses, definitely indicate that a violent explosion occurred within the tank itself and not in the surrounding spaces of the barge. The fire, which lasted several minutes after the explosion, and the blackening of parts of the interior of the tank by a slight sooty deposit, as well as the general nature of the damage indicate that the explosion was due to the ignition of some carbonaceous material (gas or vapour)'' 
\end{quote}
wrote Taylor. 
\begin{quote}
``The only hole which was burnt through the plate by the oxy-acetylene burner was a small one, which was not finished at the time of the explosion.'' ``Acetylene gas may ... be dismissed as a possible source of the explosion''; ``it must .. be concluded that the combustible matter in the tank was present when the barge arrived at the wharf'' 
\end{quote}
and 
\begin{quote}
``the whole of the circumstances of the explosion are consistent with the theory that it was caused by the ignition and subsequent explosion of a mixture of petrol vapour and air in the tank.'' 
\end{quote}

Taylor pointed out that ``no test of the atmosphere of the tank for petrol was made after the steaming''. He then performed the following calculation:
\begin{quote}
``The quantity of petrol required to render the atmosphere of the tank explosive can be readily ascertained. The tank has a capacity of 7,040 cubic feet and the vapour of petrol of the quality carried in the tank is explosive when mixed with air in proportions from 1 1/2 to 6 per cent., and about 3 per cent.\ of vapour or 210 cubic feet would produce a violently explosive mixture in the tank. One cubic foot of petrol produces about 147 cubic feet of vapour at normal temperature and pressure, so that 1 1/2 cubic feet of petrol would produce over 210 cubic feet of vapour. The total area of the interior surface of the tank and bulkheads is over 3,200 square feet, so that an extremely thin film of petrol remaining on the whole interior surface would produce explosive conditions when vapourised. Even a film less than 1/450th of an inch in thickness on the bottom alone would be sufficient to produce such conditions, whilst 1 1/2 cubic feet of liquid could easily be disposed in and around the various angle and plate joints''. 
\end{quote}

Taylor's findings were clear and unequivocal:
\begin{quote}
``Having reached the conclusion that the tank contained a mixture of petrol vapour and air on its arrival at Messrs.\ Hodge's wharf, the source of the ignition is not difficult to trace in the light of subsequent events. It is clear from the evidence of several witnesses that an oxy-acetylene burner was used to remove rivet heads on the outside of the tank, and that just when the burner was being used to burn a hole in the plate of the tank top, the explosion occurred. An examination of the damaged tank confirms this evidence in every respect and shows that the explosion took place as soon as the flame of the burner penetrated the plate.'' 
\end{quote}

Taylor laid the responsibility with the barge owners: 
\begin{quote}
``I consider that adequate precautions were not taken by the owners of the barge to ensure that the tank was free from petrol vapour before handing her over for repairs''.
\end{quote}
It turned out, Taylor found, that other barges of the same type had previously been taken to Hodge's for the same modifications to be carried out, after the same cleaning procedure, and had not exploded. The difference in those previous cases was that the barges had lain a day at the wharf before being worked on. That must inadvertently have saved the men who had worked on them, as the remaining liquid would have have an extra day to evaporate, and the vapours would 
have had an extra day to disperse before work began. 

G. Stevenson Taylor was the first Senior Engineering Inspector in the newly formed Engineering Branch of the inspectorate of factories for the Home Office. Taylor had been appointed in April 1920 and the Warwick explosion was his first report in this capacity \cite{crooks2012}.
Around this time there was a new drive for health and safety at work, both in Britain and beyond, with the slogan ``safety first''. In 1917 the Industrial ``Safety First'' Committee was established in Britain, and in 1918 the British Industrial Safety First Association was formed \cite{djang1942}. Harold Lloyd's 1923 film ``Safety Last!'', famous for the scene of Lloyd dangling from the hands of a clock high on a skyscraper, played with the slogan. Ironically, Lloyd himself had lost a finger and thumb in an explosion: a 1919 filming accident with a fake bomb that wasn't \cite{lloyd2009}.

\section{Post 1920}

Despite \emph{Safety First}, despite the advance of both science and technology, an inflammable vapour-- or gas--air mixture plus a source of ignition  continued to cause accidents.
After 1920, such explosions have continued, even into recent decades. 
Just a few of these, that illustrate the variety of detailed causes of such accidents  and their societal implications, are the following.  
The sewers of Cleveland, Ohio, exploded on 20th October 1944 when methane from a storage tank leaked into them and was ignited. The resulting explosions and fires killed 130 people and destroyed a large area of the city \cite{rydbom2013}. 
On 7th September 1951, a storage tank exploded at the Royal Edward Dock, Avonmouth, Bristol, while being filled with gas oil, and killed two men \cite{watts1952}. The Apollo 13 mission of April 1970, meant to be the third space mission to land on the moon, was aborted half way to the moon, and its crew was narrowly saved after an explosion aboard the spacecraft that occurred when stirring an oxygen tank for its fuel cells, which had been incorrectly assembled leaving damaged electrical insulation on the wiring, which ignited \cite{cortright1970}. The Clarkston explosion of October 1971 killed 22 people at a shopping centre in Scotland when a gas leak  ignited. The Flixborough disaster of 1974 was an explosion at a chemical works that killed 28 people and ``rattled the confidence of every chemical engineer in the country'' \cite{kinnersley1975} when a pipe joint in a poorly modified cyclohexane plant failed and the resulting gas cloud was ignited \cite{parker1975,newland1976,davidson2020}.
A fuel tanker explosion of July 1978 in Los Alfaques, near Tarragona, Spain, killed 217 people when a tanker lorry carrying propene leaked gas which exploded on encountering an ignition source in a neighbouring seaside campsite.
\emph{Piper Alpha} was an oil rig in the North Sea that exploded in July 1988 killing 167 people working on it when methane hydrate blocked a pump and another pump, partially dismantled for maintenance, leaked inflammable gas when mistakenly switched on to take its place \cite{cullen1990}.
Flight TWA800 was a Boeing 747 aircraft that exploded in July 1996 minutes after taking off from New York killing its 230 passengers and crew. Its loss was determined to be owing to an explosion in its empty central fuel tank in which inflammable vapour was ignited by a short circuit in the electrical connections \cite{ntsb2000}.
The Buncefield accident of December 2005 at an oil storage facility in Hertfordshire killed no-one but caused an immense explosion and fire when a petrol tank was overfilled and a cloud of vapour from it was ignited \cite{hse2008}.  The \emph{Deepwater Horizon} was an oil rig in the Gulf of Mexico that suffered a blowout of inflammable gas that was ignited probably by coming into contact with the diesel generators on board \cite{bp2010}; 11 people died in that April 2010 explosion, and it produced a huge oil spill that caused environmental devastation in the Gulf of Mexico.

\section{The theory of fluid mechanics of ignition, combustion, and explosion}

An explosion is a rapid increase in volume. Accompanying this, there are pressure waves and the loud bang we associate with an explosion. Whether or not a chemical reaction in a fluid leads to an explosion depends on the interaction between fluid mechanics and 
chemical reaction. 
We may call these explosions thermo-kinetic because, as we shall see, the explosion --- i.e., fast reaction --- can result from self-heating (thermal) or from the chemical kinetics (chemical chain reaction) or from both factors. 

Combustion and explosion science \cite{lewis1987,nettleton1987} is now a field with some beautiful and elegant theory \cite{zeldovich1985,buckmaster2005}, and experiments \cite{oppenheim1973}.
However, not everything is known. There are still areas in which more knowledge of the processes taking place is needed  \cite{babrauskas2007}, as we shall discuss.
It must also be noted that, as we have seen, the impact of thermal radiation from these explosions is a major factor in terms of human health, and this aspect is also the object of study \cite{hymes1996,beyler2016}.
Combustion theory began to be developed in 1880s France where Mallard and Le Chatelier worked on inflammability of gases; combustion and its application to mine safety \cite{manson1988}. They developed a laminar flame speed theory for the 
rate of expansion of the flame front in a combustion reaction  \cite{mallard1880,mallard1883}.
At almost the same time, van't Hoff worked on autoignition of gases and suggested a criterion for when ignition would occur  \cite{vanthoff1884}.

Semenov and his student  Frank-Kamenetskii took up these ideas in 1930s Russia. Semenov's theory \cite{semenov1928,semenov1959} pertains to the case where  convection is so vigorous that the temperature in the vessel is uniform; Frank-Kamenetskii's \cite{frank1938,frank1969} covers the other limit where the transport of heat occurs by conduction only. 
For well-mixed systems, the reacting gas explodes when the Semenov number exceeds a critical value, when $$\psi=k_0c_0^nqE/(\chi S_vRT_0^2)$$ is greater than $1/e$. (Here $k_0$ is the initial kinetic constant for first-order reaction A $\to$ B, $c_o$ is  the initial concentration of species A, $n$ is the order of the reaction, $q$ is the reaction exothermicity, $E$ is the activation energy, $\chi$ is the heat transfer coefficient, $S_v$ is the surface area per unit volume, $R$ is the universal gas constant, $T_0$ is the constant wall temperature.)
 When heat is transferred by thermal conduction alone, explosion occurs when the Frank-Kamenetskii number is greater than a critical value, when $$\delta=l^2k_0c_0^nqE/(\kappa \rho_0 C_pRT_0^2)$$ is greater than $\delta_{c}$. (Here $l$ is the reactor size, $k_0$ is the initial kinetic constant for first-order reaction A $\to$ B, $c_0$ is the initial concentration of species A, $n$ is the reaction order, $q$ is the reaction exothermicity, $E$ is the activation energy, $\kappa$ is the thermal diffusivity, $\rho_0$ is the initial density, $C_p$ is the specific heat at constant pressure, $R$ is the universal gas constant, $T_0$ is the constant wall temperature.
 The critical value depends on the geometry of the system; it is 3.32 for a sphere, 2.00 for a cylinder and 0.88 for parallel plates \cite{frank1969}.)
But in most reacting systems heat loss occurs due to the combined effects of natural convection and heat conduction. 
In the 1950s, Kistiakowsky and coworkers published a series of experimental papers with the common title  \emph{gaseous detonations} \cite{berets1950,berets1950_2,kistiakowsky1952,kistiakowsky1952_2,kistiakowsky1952_3,kistiakowsky1955,kistiakowsky1955_2,kistiakowsky1956,kistiakowsky1956_2,chesick1958,cher1958,kistiakowsky1959,kistiakowsky1961}. Among other aspects of the conditions for explosion that they investigated was the role in explosion of temperature changes  due to heat conduction. Sokolik's 1960 textbook \emph{Self-ignition, Flame and Detonation in Gases}  (published in English in 1963) \cite{sokolik1963} models these conditions of explosion and the role of the non-dimensional activation energy term in the peaking of heat release, crucial for any explosion.

In parallel to these works, fluid mechanics was developing. The equations for the movement of a fluid were written down in the early 19th century by Navier, Cauchy, Poisson,  Saint Venant, and Stokes \cite{stokes}. In the second half of the 20th century, with the advent of the electronic computer it became possible to solve the Navier--Stokes equations numerically. Computational fluid dynamics, CFD, methods are now ubiquitous. One can couple these computational methods with the equations of chemical kinetics to solve numerically how a reaction behaves in a fluid medium. 
Three physical processes combine in a reacting flow:  fluid dynamics, thermodynamics, and chemical reaction. 
Each process has its own space- and time-scales, which may be very different from those of the other processes. Such differences of scales, on one hand, can allow simplification of a theoretical model, but on the other hand, they can also be a source of numerical difficulties. The fluid dynamics is a balance between the temporal evolution and the spatial convection of the flow properties governed by the conservation of mass, momentum and energy. Reactive fluid thermodynamics includes microscopic heat transfer between gas molecules, work performed with pressure and associated volume change. And chemical reactions determine the generation and  destruction of chemical species while observing mass conservation.

One application of the theory of chemically reactive flows is to mining, as its pioneers had studied. We may note that both original types of miners' safety lamps work because the mesh or tubes function as a flame arrestor. The metal absorbs heat from a flame front, so that the front decays and the flame dies. This mechanism depends critically on the size of the tubes or the mesh. 
And, as we have seen, Faraday commented on the choice before safety lamps of a small candle --- one presumably having a small flame --- by miners, which might provide illumination without igniting the firedamp.

Work on strong and weak ignition can be mentioned at this point. 
In the 1960s, Voevodsky and Soloukhin looked at variations in ignition arising from  chemical kinetics, and defined two regimes \cite{voevodsky1965}. Strong ignition occurs when a single ignition point is enough to give a detonation front. Weak ignition takes place at lower temperature when reaction is initiated at multiple points. These coalesce to produce a front. 
Meyer and Oppenheim showed that weak ignition occurs at different points where fluid flow velocities are low, and that the induction time to reaction is sensitive to temperature variations owing to inhomogeneities \cite{meyer1971}.  This understanding of  the different processes of weak and strong ignition is of relevance to the deflagration-to-detonation transition that we shall discuss below
\cite{boeck2017}.

As well as mining, a further application, this time one where explosions are wanted, but need to be controlled, is the internal combustion engine.  An amount of work has been carried out under the heading of knock, or preignition, in the internal combustion engine \cite{willand1998,lewis1987}. (It should be noted that this is generally a two phase system, as droplets of fuel, not vapour are the reactant. It has been argued that fuel involved in the Buncefield explosion, described above, may have been in the form of such a mist of droplets  \cite{taveau2012}.)
Knock in internal combustion engines has been associated with cool flame formation, and  cool flames resulting from complex thermo-kinetic interactions \cite{campbell2008} in turn have been implicated in the TWA800 explosion mentioned above.

\begin{figure}[tbp]
\centering\includegraphics*[width=0.7\columnwidth,clip=true]{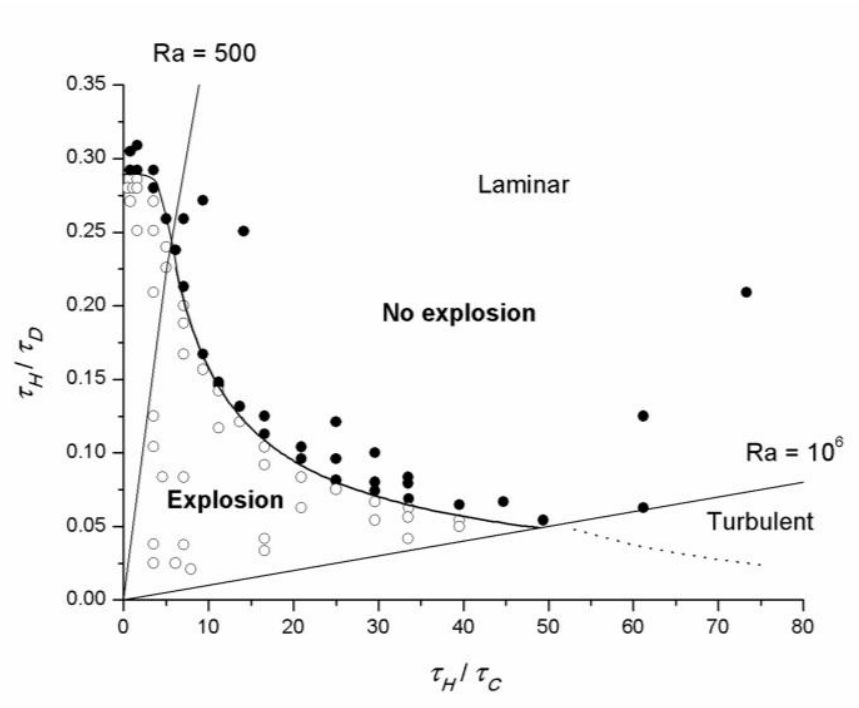}
\caption{\label{fig:2Dtimescales}
The regime diagram summarizing simulations without consumption of reactant, with closed circles representing no explosion and open circles, explosion. $\tau_H$ is the timescale for  heating by reaction; $\tau_C$ is  the timescale for  convection, $\tau_D$ is the timescale for diffusion. The horizontal axis then denotes the well-mixed limit; the vertical axis represents the purely conductive limit. 
The relative importance of thermal conduction and natural convection in the system is shown by the Rayleigh number $Ra$;
when $Ra < 500$, heat transfer is controlled by conduction; laminar convection dominates heat transfer for $500<Ra<10^6$; for $Ra > 10^6$, the flow is turbulent.
The other solid and dotted lines, respectively, represent the explosion limit in the laminar and turbulent regimes. From  \cite{liu2008} with permission from the PCCP Owner Societies.
}
\end{figure}

Recent theoretical work in the last two decades, which takes the Semenov and Frank-Kamenetskii approaches and moves them forward, is by the group of Cardoso in Cambridge. Following an approach first put forward by Cardoso et al. \cite{cardoso2004_1,cardoso2004_2},
Liu et al.\ \cite{liu2008} proposed that in these systems the occurrence or not of an explosion depends on the relative magnitudes of three timescales: that for chemical reaction to heat up the fluid to ignition, the timescale for thermal conduction and that for natural convection. They summarized their results in a two-dimensional regime diagram, Fig.~\ref{fig:2Dtimescales}, in which Frank-Kamenetskii's purely conductive system and Semenov's well-mixed system appear as two limiting cases, represented by the two axes. The plane in between the two axes contains all the systems with different relative magnitudes of heat loss by conduction and by natural convection. This approach has the advantage of quantifying separately the stabilizing effects of conduction and natural convection on an explosion.

In a subsequent paper, they then extended the these ideas to explore how the consumption of reactant alters the onset of a thermal explosion \cite{liu2010}. They showed that 
whether or not a chemical reaction in a fluid leads to an explosion is shown to depend on four timescales: (1)
that for the chemical reaction to heat up the fluid containing the reactants and products, 
$$\tau_H=\frac{\rho_0 C_p \Delta T_s}{k_0 c_0^nq},$$
$\rho_0$ being the initial density, $C_p$  the specific heat at constant pressure, $k_0$ is the initial kinetic constant for first-order reaction A $\to$ B, $c_o$ is  the initial concentration of species A, $n$ is the order of the reaction, $q$ is the reaction exothermicity, and $\Delta T_s = RT_0^2/E$ ($E$ being an activation energy) is a temperature increase scale; (2)
for cooling by heat conduction or diffusion out of the system, 
$$\tau_D=\frac{l^2}{\kappa},$$ 
$l$ being the reactor size and $\kappa$ the thermal diffusivity; 
(3) for natural convection in the fluid, 
$$\tau_C=\frac{l}{U},$$
$U$ being a characteristic velocity; 
and finally (4) for chemical reaction that uses up the reactant. 
$$\tau_R=1/k_0,$$ 
$k_0$ being the initial kinetic constant for first-order reaction A $\to$ B.

\begin{figure}[tbp]
\centering\includegraphics*[width=0.7\columnwidth,clip=true]{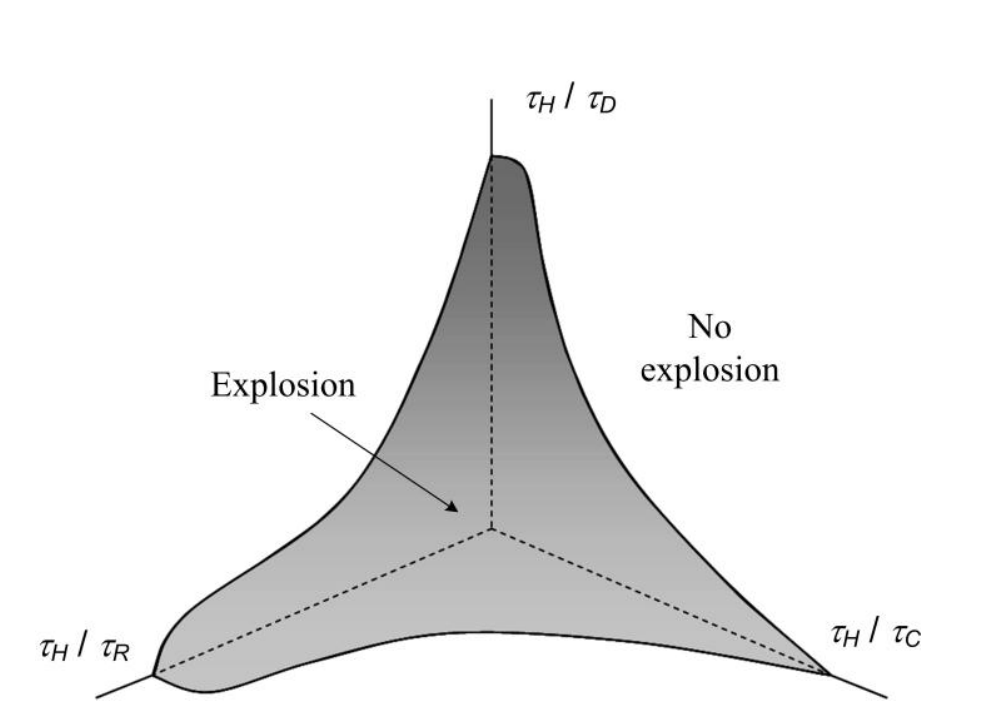}
\caption{\label{fig:timescales}
Schematic regime diagram with the three axes, $\tau_H /\tau_C$, $\tau_H/\tau_D$ and $\tau_H / \tau_R$, for an exothermic
reaction occurring inside a closed spherical vessel.  $\tau_H$ is the timescale for  heating by reaction; $\tau_C$ is  the timescale for  convection, $\tau_D$ is the timescale for diffusion and $\tau_R$ is the timescale for reaction. The grey surface separates the inner region, where explosions occur, from the outer one, where they do not. From  \cite{liu2010} with permission of Elsevier.
}
\end{figure}

The behaviour of the system can be depicted on a three-dimensional regime diagram, as shown in Fig.~\ref{fig:timescales}, where the three ratios of the four timescales are the coordinates. 
There is a surface separating the region near the origin where explosions occur from the region further from the origin where the stabilizing effects of heat conduction, natural convection and consumption of reactant prevent explosions. The vertical axis $\tau_H/\tau_D$ represents the purely conductive limit ignoring depletion of reactant, i.e., the systems considered by Frank-Kamenetskii. The right-hand axis, $\tau_H /\tau_C$ gives the well-mixed limit of Semenov with no consumption
of reactant. And the third, left-hand axis $\tau_H / \tau_R$ measures the effect of the disappearance of reactant on the heating up of the fluid; for example, if the chemical reaction in effect depletes the reactant much faster than fluid is heating up (e.g., because the heat of reaction is very small), then $\tau_H  \gg \tau_R$ and we expect the temperature rise in the fluid to be small and explosion not to occur.
 The synthesis of these two limits into one --- thermal and kinetic effects into a thermo-kinetic theory --- is a beautiful piece of theory,  bridging and bringing together points often seen as separate, but which really are not separate at all, as the figures show.
 
 \begin{figure}[tbp]
\centering\includegraphics*[width=0.9\columnwidth,clip=true]{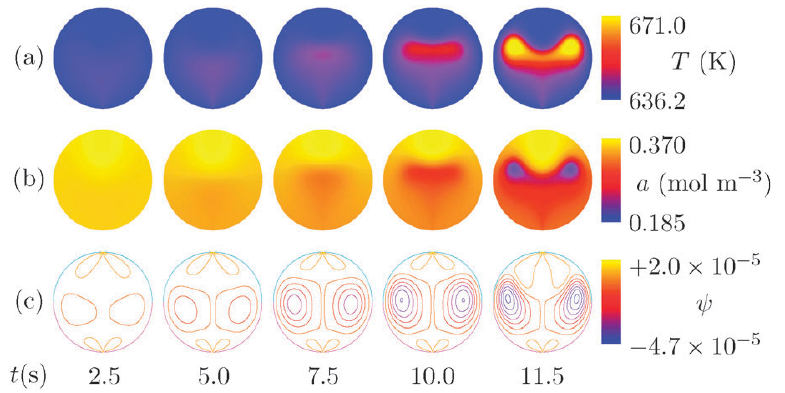}
\caption{\label{fig:sim_explosion}
Numerical simulations with computational fluid dynamics of explosion in a spherical reactor with mixed convection; evolution of the fields of (a) temperature, (b) reactant concentration and (c) streamlines.
The temperature starts rising at the bottom of the vessel, accompanied by reactant depletion, and buoyancy is triggered within the fluid. Owing to the interaction between buoyancy and forced convection the hot region expands further to both sides of the reactor where the cooling by diffusion of colder species from the inlet stream is not so efficient. This allows the fluid to heat up by reaction, leading to explosion at $t = 11.5$~s. From \cite{azevedo2014} with permission from the PCCP Owner Societies.
}
\end{figure}

The Cardoso group also performed numerical CFD simulations of the effects of combined natural and forced convection on thermal explosion in a spherical reactor with upward natural convection from internal heating caused by a chemical reaction to which they added  downward forced convection driven by injecting fluid at the top and removing it at the bottom of the reactor \cite{liu2013,azevedo2014}. They  found oscillatory behaviour for moderate forced convection. They also found that that explosive behaviour is favoured by a balance between the natural and forced flows, owing to a nearly stagnant zone close to the centre of the reactor that quickly heats up to explosion; Fig.~\ref{fig:sim_explosion}. It is counter-intuitive that explosion may occur in an otherwise stable reactor by injecting cold fluid or enhancing natural convection.

The foregoing theory is for a homogeneous mixture of reactants in a  vessel with its walls at constant temperature. One significant aspect where theory has been inadequate and computational fluid dynamics needed is the deflagration to detonation transition \cite{oran}, important in the context of the confinement of an explosion.
The central theme of this review is to illustrate the consequences of pressure build up from the increase of volume in an unventilated medium. The pressure waves and loud bang of an explosion have not been discussed, because our focus has been not on the waves and bang but on the time scales of the phenomenon in the context of the fluid mechanics and the chemistry leading to volume increase. 
Quasi-detonations and detonations and their initiation, either through transition from a flame or otherwise have a strong gas dynamic component. Quasi detonations are more likely in an explosion.
While it is not possible to do justice to these other aspects of explosions here, for which we send the reader to recent reviews \cite{kundu2016,wang2017,chamberlain2019,xu2020,oran2020}, it is worth mentioning them because they show how science always needs to be interdisciplinary, because it turns out that the same science is involved in the explosions of industrial accidents and in the explosions of stars. 

\begin{figure}[tbp]
\centering\includegraphics*[width=\columnwidth,clip=true]{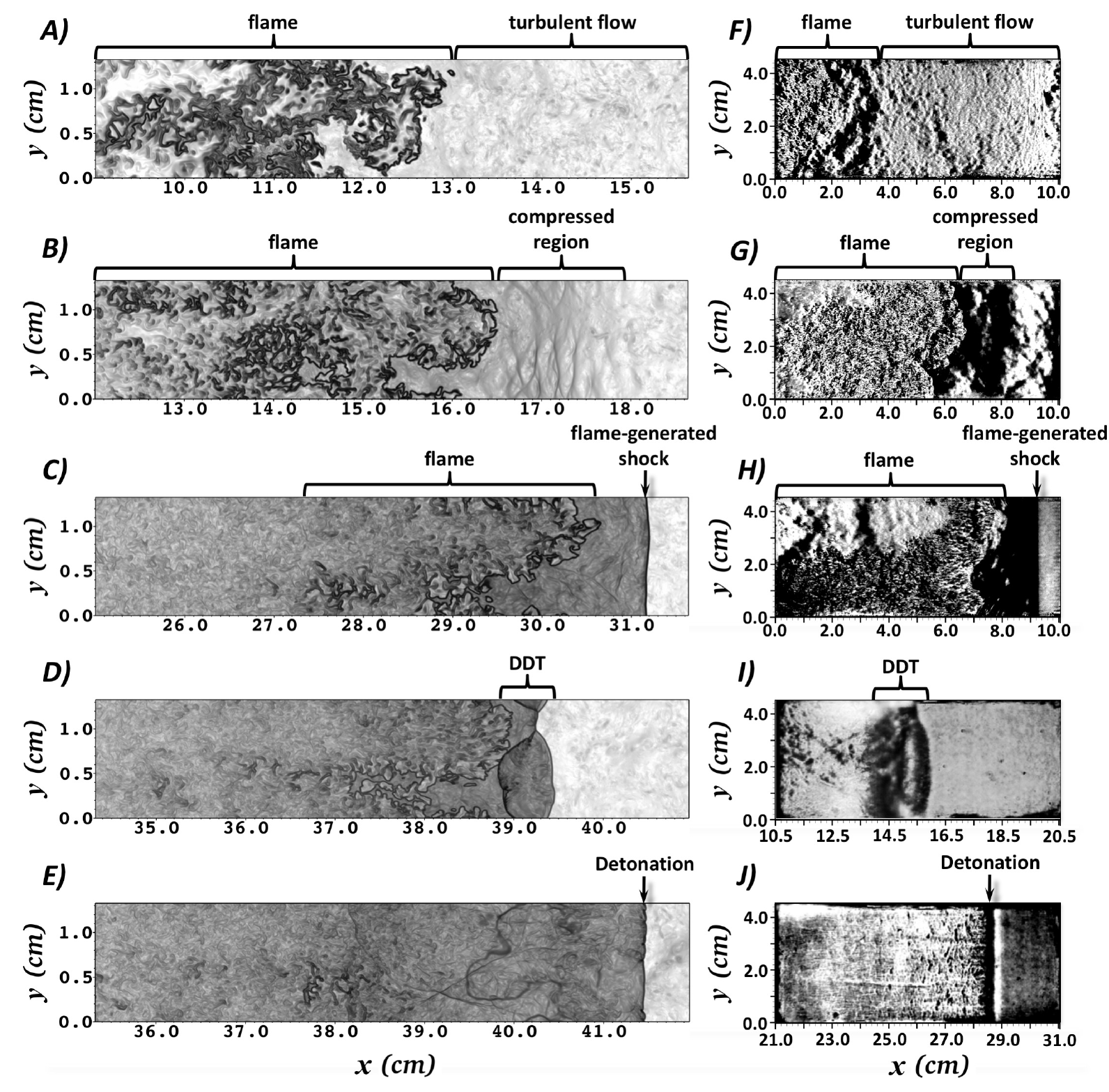}
\caption{\label{fig:schlieren}
Simulated (A--E) and experimental (F--J) schlieren images of turbulent flame evolution.
Once the flame front has developed, it begins to propagate toward the open end of the channel (A and F).
Pressure waves generated within the turbulent flame propagate into the unburned material and form a compressed region ahead of the flame (B and G). As the runaway process develops, multiple pressure waves coalesce into a flame-generated shock, the strength of which grows with time (C and H). Eventually, the shock triggers a deflagration-to-detonation transition, DDT (D and I) and gives rise to detonation (E and J). From \cite{poludnenko2019}. Reprinted with permission from AAAS.
}
\end{figure}

 In deflagration there is a subsonic flame propagation velocity and in the reaction zone, chemical combustion progresses through the medium by the diffusion of heat and mass. In contrast, in a detonation there is a supersonic flame propagation velocity, and the reaction zone is a shock wave where the reaction is initiated by  compressive heating caused by the shock wave. This regime has generally needed numerical simulations of computational fluid dynamics for its understanding. 
 A general theory of the deflagration-to-detonation transition has recently been presented together with an interesting astrophysical application to the explosion of stars. 
Figure~\ref{fig:schlieren} shows schlieren images with the coalescing of pressure waves and formation of a shock wave.
Poludnenko et al show that thermonuclear combustion waves in type Ia supernovae are qualitatively similar to chemical combustion waves on Earth because they are controlled by the same physical mechanisms and are not sensitive to the details of the equation-of-state, microphysical transport, or reaction kinetics \cite{poludnenko2019}.  They discuss how a deflagration-to-detonation transition may have been involved in some industrial accidents, such as at Buncefield. 
At the same time, the astrophysical application demonstrates how interdisciplinary this field at the interface between fluid mechanics and chemistry is, and how far from terrestrial concerns it can get.

\section{Safety first?}

During the past century, combustion theory has grown from its incipient state to fully fledged. The understanding of chemical reaction kinetics has evolved immensely.  Fluid mechanics has developed both
theory and also numerical solutions with CFD methods as computers  first appeared and subsequently have doubled in speed every eighteen months or so. Engineering science --- chemical, but also mechanical, civil, and other branches --- has likewise advanced at an ever increasing pace. And health and safety planning has progressed enormously in foreseeing and eliminating risks since the early days of factory and engineering inspectors and \emph{safety first}. Nonetheless, gas and vapour explosions have killed many people, and in many of these cases a container or tank was involved, as in the accident of my grandfather. 

In the last 100 years, there have been developed and implemented systems for making safer such (especially) fuel tanks with a nitrogen or similar inert atmosphere, or a non-combustible atmosphere such as carbon dioxide.
Some second world war aircraft used inerting systems, and tanker ships trialled them in the 1920s.
On-board inert gas generation systems have been developed for aircraft \cite{langton2010}.
Tanker ships use either gases produced in combustion by the ship engines, or they generate inert gases \cite{nmab1980}.
Likewise, so-called hypoxic or oxygen-reduced air systems with low-oxygen-concentration air have been developed for buildings: for archives, warehouses, electrical substations and other buildings where reducing  is preferred to eliminating oxygen \cite{nilsson2013}.
And racing-car fuel tank or fuel cell technology involves filling the tank with an open-cell foam, again to reduce the explosion risk.
We can see all these safety systems as playing with the parameters of the theory that we have described above, to move from an explosive to a non-explosive part of the parameter space. So why have such safety systems not, like miners' safety lamps, become the norm? It is arguable that in some instances,  that of cars, for instance, car fuel tank explosions are (despite the mythopoeia of Hollywood) rather rare. This does not appear to be the case, however, for tanker ships.
Devanney \cite{devanney2010}  compiled a long list of tanker ship accidents caused by explosions in fuel tanks.
He asks regarding inerting, 
\begin{quote}
``why was the industry so slow to adopt such an obvious, effective, safety measure which probably pays for itself in reduction of tank corrosion?'' 
\end{quote}
and answers his own question
\begin{quote}
``the cost[s] of implementing inerting to these ships were more than the dollar benefits of the lives and ships saved'', 
\end{quote}
because 
\begin{quote}
``a tanker owner rarely suffers any loss when one of his ships blows up. His P\&I [Protection and Indemnity] insurance pays off the dead crew's family, in most past cases a few thousand dollars per head. His hull insurance covers the loss of the ship. In many cases, the insured value is more than the market value and the owner comes out ahead.''
\end{quote}

This review has demonstrated how, from the very beginning, the science of explosions has been inextricably intertwined with its societal implications. 
As I write, at  the centenary of my grandfather's death, 
I do wonder whether more could not be done, with more innovative solutions from  scientific and technological ideas, to reduce further the risk of explosion, in the same way that miners' safety lamps were developed and used.
Despite great advances in both science and engineering, gas and vapour explosions continue \cite{atkinson2017}. 
I began this review with  sewer explosions of two millennia ago and I have mentioned the enormous 1944 disaster in Cleveland, Ohio; these still  occur on occasion.
A fuel tanker blew up a century ago owing to incorrect working methods with fuel vapour. That was then, but how can it be that a century on fuel tankers are still blowing up for the same reason? Has it really  been \emph{safety first}, or \emph{safety last}?

Whether or not further regulatory action is needed is a question that scientists involved with the societal responsibilities of science should consider.  Some scientists believe that we should not concern ourselves with these matters, which they see as 
 political rather than scientific. To my mind that is a fundamentally misguided point of view.
I myself firmly believe that we as scientists must be open to, and understand, the societal aspects of science, that science owes it to society to explain the implications of science for society, and these matters should be discussed alongside the science itself.
From the point of view of the science itself, in this field science has always engaged with the practical questions arising from explosions.  The Felling pit explosion of 1812 led to the miners' safety lamp; the Deepwater Horizon explosion of 2010 and the resulting oil spill in the waters below the rig has led to a great quantity of research into the dynamics of 
two-phase plumes \cite{domingos2013}, and the Buncefield explosion of 2005 is noted in the latest results on supernovae explosions \cite{poludnenko2019},  I am sure that new scientific advances will continue to emerge from the fluid mechanics of explosion.

\section*{Acknowledgements}
I thank Silvana Cardoso and John Davidson (1926--2019) for many interesting discussions  in the tea room of the Department of Chemical Engineering and Biotechnology in Cambridge over the years that have contributed to this review,  ranging from the fluid mechanics of explosions, to the lessons of Flixborough for the education of engineers, to how Tom Bacon  developed, first at  C. A. Parsons in Newcastle and then in that Department, the hydrogen--oxygen fuel cell used by NASA in the Apollo programme.
I acknowledge the financial support of the Spanish MINCINN project FIS2016-77692-C2-2-P.

\bibliographystyle{unsrt}
\bibliography{barge}

\begin{thebibliography}{10}

\bibitem{scobie1986}
A.~Scobie.
\newblock Slums, sanitation and mortality in the {Roman} world.
\newblock {\em Klio}, 68(2):399, 1986.

\bibitem{koloski2015}
A.~O. Koloski-Ostrow.
\newblock {\em The Archaeology of Sanitation in Roman Italy: Toilets, Sewers,
  and Water Systems}.
\newblock The University of North Carolina Press, 2015.

\bibitem{deming2020}
D.~Deming.
\newblock The aqueducts and water supply of ancient {Rome}.
\newblock {\em Groundwater}, 58:152--161, 2020.

\bibitem{morgan1936}
J.~R. Morgan.
\newblock The search for a safety-lamp in mines.
\newblock {\em Annals of Science}, 1:302--329, 1936.

\bibitem{davy1816}
H.~Davy.
\newblock On the fire-damp of coal mines, and on methods of lighting the mines
  so as to prevent its explosion.
\newblock {\em Phil. Trans. Roy. Soc.}, 106:1--22, 1816.

\bibitem{brandling1817}
C.~J. Brandling.
\newblock {\em Report upon the claims of Mr. George Stephenson, relative to the
  invention of his safety lamp, by the committee appointed at a meeting holden
  in Newcastle on the First of November 1817}.
\newblock Emerson Charnley, 1817.

\bibitem{faraday}
M.~Faraday.
\newblock {\em The chemical history of a candle}.
\newblock Griffin, Bohn and Company, London, 1861.

\bibitem{thomas2015}
J.~M. Thomas.
\newblock {Sir Humphry Davy} and the coal miners of the world: A commentary on
  {Davy} (1816) `{An} account of an invention for giving light in explosive
  mixtures of fire-damp in coal mines'.
\newblock {\em Phil. Trans. Roy. Soc. A}, 373:20140288, 2015.

\bibitem{Note1}
There is no difference between a gas and a vapour in terms of their fluid
  physics or chemistry; only in terms of their thermodynamics. The different
  terms simply indicate that a gas is a substance found at that temperature and
  pressure only in the single thermodynamic gaseous state, whereas a vapour is
  also found in its condensed phase at the same temperature and pressure.

\bibitem{graham1853}
T.~Graham.
\newblock Chemical report on the cause of the fire in the {Amazon}.
\newblock {\em Q. J. Chem. Soc. London}, 5:34--96, 1853.

\bibitem{taylor1874}
A.~S. Taylor.
\newblock The {Regent's Park} explosion.
\newblock {\em British Med. J.}, 2:477, 1874.

\bibitem{holland1874}
E.~Holland.
\newblock The {Regent's Park} explosion.
\newblock {\em British Med. J.}, 2:478, 1874.

\bibitem{abel1875}
F.~A. Abel.
\newblock Accidental explosions.
\newblock {\em Nature}, 12:436--439, 477--478, 498--499, 1875.

\bibitem{abel1885}
F.~A. Abel.
\newblock Accidental explosions produced by non-explosive liquids.
\newblock {\em Nature}, 31:469--472, 493--496, 518--521, 1885.

\bibitem{redwood1894}
B.~Redwood.
\newblock The transport of petroleum in bulk.
\newblock In {\em Minutes of the Proceedings of the Institution of Civil
  Engineers}, volume 116, pages 177--229, 1894.

\bibitem{taylor1921}
G.~S. Taylor.
\newblock {\em Report on an Explosion on the Oil Tank Barge ``{Warwick}'' on
  {September} 24th, 1920}.
\newblock HMSO, London, 1921.

\bibitem{crooks2012}
E.~Crooks.
\newblock {\em The Unrelenting Machine: A legacy of the industrial revolution}.
\newblock lulu.com, 2012.

\bibitem{djang1942}
T.~K. Djang.
\newblock {\em Factory Inspection in Great Britain}.
\newblock George Allen \& Unwin, 1942.

\bibitem{lloyd2009}
A.~D'Agostino Lloyd.
\newblock {\em Harold Lloyd: Magic in a Pair of Horn-Rimmed Glasses}.
\newblock BearManor Media, Albany, GA, 2009.

\bibitem{rydbom2013}
C.~Rydbom and T.~Kubat.
\newblock {\em Cleveland Area Disasters}.
\newblock Arcadia Publishing, Charleston, South Carolina, 2013.

\bibitem{watts1952}
H.~E. Watts.
\newblock {\em Report on Explosion and Fire at {Regent Oil Co. Ltd.} premises
  {Avonmouth, Bristol} on 7th {September} 1951}.
\newblock HMSO, London, 1952.

\bibitem{cortright1970}
E.~M. Cortright.
\newblock {\em Report of Apollo 13 Review Board}.
\newblock NASA, 1970.

\bibitem{kinnersley1975}
P.~Kinnersley.
\newblock What really happened at {Flixborough}?
\newblock {\em New Scientist}, 65:520--522, 1975.

\bibitem{parker1975}
R.~J. Parker, J.~A. Pope, J.~F. Davidson, and W.~J. Simpson.
\newblock {\em The Flixborough Disaster --- Report of the Court of Inquiry}.
\newblock HMSO, London, 1975.

\bibitem{newland1976}
D.~E. Newland.
\newblock Buckling and rupture of the double bellows expansion joint assembly
  at {Flixborough}.
\newblock {\em Proc. Roy. Soc. A}, 351:525--549, 1976.

\bibitem{davidson2020}
J.~F. Davidson.
\newblock Life and times in engineering and chemical engineering.
\newblock {\em Annu. Rev. Chem. Biomol. Eng.}, 11:23--34, 2020.

\bibitem{cullen1990}
W.~D. Cullen.
\newblock {\em The public inquiry into the {Piper Alpha} disaster}.
\newblock HMSO, London, 1990.

\bibitem{ntsb2000}
{\em In-flight Breakup Over the Atlantic Ocean Trans World Airlines Flight 800
  Boeing 747-131, N93119 Near East Moriches, New York July 17, 1996}.
\newblock National Transportation Safety Board, 2000.

\bibitem{hse2008}
{\em The Buncefield Incident 11 December 2005: The final report of the Major
  Incident Investigation Board}.
\newblock HSE Books, 2008.

\bibitem{bp2010}
{\em {Deepwater Horizon} Accident Investigation Report}.
\newblock BP, 2010.

\bibitem{lewis1987}
B.~Lewis and G.~von Elbe.
\newblock {\em Combustion, Flames and Explosions of Gases}.
\newblock Academic Press, third edition, 1987.

\bibitem{nettleton1987}
M.~A. Nettleton.
\newblock {\em Gaseous Detonations: Their nature, effects and control}.
\newblock Chapman and Hall, 1987.

\bibitem{zeldovich1985}
Y.~B. Zeldovich, G.~I. Barenblatt, V.~B. Librovic, and G.~M. Makhviladze.
\newblock {\em The Mathematical Theory of Combustion and Explosion}.
\newblock Consultants Bureau, New York, 1985.

\bibitem{buckmaster2005}
J.~Buckmaster, P.~Clavin, A.~Linan, M.~Matalon, N.~Peters, G.~Sivashinsky, and
  F.~A. Williams.
\newblock Combustion theory and modeling.
\newblock {\em Proceedings of the Combustion Institute}, 30:1--19, 2005.

\bibitem{oppenheim1973}
A.~K. Oppenheim and R.~I. Soloukhin.
\newblock Experiments in gasdynamics of explosions.
\newblock {\em Annu. Rev. Fluid Mech.}, 5:31--58, 1973.

\bibitem{babrauskas2007}
V.~Babrauskas.
\newblock Ignition: A century of research and an assessment of our current
  status.
\newblock {\em J. Fire Protection Eng.}, 17:165, 2007.

\bibitem{hymes1996}
I.~Hymes, W.~Boydell, and B.~Prescott.
\newblock {\em Thermal radiation: Physiological and pathological effects}.
\newblock Institution of Chemical Engineers, 1996.

\bibitem{beyler2016}
C.~L. Beyler.
\newblock Fire hazard calculations for large, open hydrocarbon fires.
\newblock In M.~J. Hurley, D.~Gottuk, J.~R. Hall~Jr., K.~Harada, E.~Kuligowski,
  M.~Puchovsky, J.~Torero, J.~M. Watts~Jr., and C.~Wieczorek, editors, {\em
  SFPE Handbook of Fire Protection Engineering}. Springer, 2016.

\bibitem{manson1988}
N.~Manson.
\newblock Some notes on the first theories of the flame velocity in gaseous
  mixtures.
\newblock {\em Combustion and Flame}, 71:179--187, 1988.

\bibitem{mallard1880}
F.~E. Mallard and H.~Le~Chatelier.
\newblock Sur les temp{\'e}ratures d'inflammation des m{\'e}langes gazeux.
\newblock {\em C. R. Acad{\'e}mie Sciences Paris}, 91:825--828, 1880.

\bibitem{mallard1883}
F.~E. Mallard and H.~Le~Chatelier.
\newblock {\em Recherches exp\'erimentales et th\'eoriques sur la combustion
  des m\'elanges gazeux explosives}.
\newblock Dunod, 1883.

\bibitem{vanthoff1884}
J.~H. Van't~Hoff.
\newblock {\em Etudes de dynamique chimique}, volume~1.
\newblock Muller, 1884.

\bibitem{semenov1928}
N.~N. Semenov.
\newblock Zur theorie des verbrennungsprozesses.
\newblock {\em Zeitschrift f{\"u}r Physik}, 48:571--582, 1928.

\bibitem{semenov1959}
N.~N. Semenov.
\newblock {\em Some Problems in Chemical Kinetics and Reactivity}, volume~2.
\newblock Princeton University Press, 1959.

\bibitem{frank1938}
D.~A Frank-Kamenetskii.
\newblock Towards temperature distributions in a reaction vessel and the
  stationary theory of thermal explosion.
\newblock {\em Doklady Akad. Nauk SSSR}, 18:411--412, 1938.

\bibitem{frank1969}
D.~A. Frank-Kamenetskii.
\newblock {\em Diffusion and Heat Transfer in Chemical Kinetics}.
\newblock Plenum Press, second edition, 1969.

\bibitem{berets1950}
D.~J. Berets, E.~F. Greene, and G.~B. Kistiakowsky.
\newblock Gaseous detonations. {I}. stationary waves in hydrogen -- oxygen
  mixtures.
\newblock {\em J. Amer. Chem. Soc.}, 72:1080--1086, 1950.

\bibitem{berets1950_2}
D.~J. Berets, E.~F. Greene, and G.~B. Kistiakowsky.
\newblock Gaseous detonations. {II}. initiation by shock waves.
\newblock {\em Journal of the American Chemical Society}, 72:1086--1091, 1950.

\bibitem{kistiakowsky1952}
G.~B. Kistiakowsky, H.~T. Knight, and M.~E. Malin.
\newblock Gaseous detonations. {III}. dissociation energies of nitrogen and
  carbon monoxide.
\newblock {\em J. Chem. Phys.}, 20:876--883, 1952.

\bibitem{kistiakowsky1952_2}
G.~B. Kistiakowsky, H.~T. Knight, and M.~E. Malin.
\newblock Gaseous detonations. {IV}. the acetylene-oxygen mixtures.
\newblock {\em J. Chem. Phys.}, 20:884--887, 1952.

\bibitem{kistiakowsky1952_3}
G.~B. Kistiakowsky, H.~T. Knight, and M.~E. Malin.
\newblock Gaseous detonations. {V}. nonsteady waves in {CO--O}$_2$ mixtures.
\newblock {\em J. Chem. Phys.}, 20:994--1000, 1952.

\bibitem{kistiakowsky1955}
G.~B. Kistiakowsky and P.~H. Kydd.
\newblock Gaseous detonations. {VI}. the rarefaction wave.
\newblock {\em J. Chem. Phys.}, 23:271--274, 1955.

\bibitem{kistiakowsky1955_2}
G.~B. Kistiakowsky and W.~G. Zinman.
\newblock Gaseous detonations. {VII}. a study of thermodynamic equilibration in
  acetylene-oxygen waves.
\newblock {\em J. Chem. Phys.}, 23:1889--1894, 1955.

\bibitem{kistiakowsky1956}
G.~B. Kistiakowsky and P.~C. Mangelsdorf~Jr.
\newblock Gaseous detonations. {VIII}. two-stage detonations in
  acetylene-oxygen mixtures.
\newblock {\em J. Chem. Phys.}, 25:516--519, 1956.

\bibitem{kistiakowsky1956_2}
G.~B. Kistiakowsky and P.~H. Kydd.
\newblock Gaseous detonations. {IX}. a study of the reaction zone by gas
  density measurements.
\newblock {\em J. Chem. Phys.}, 25:824--835, 1956.

\bibitem{chesick1958}
J.~P. Chesick and G.~B. Kistiakowsky.
\newblock Gaseous detonations. {X}. study of reaction zones.
\newblock {\em J. Chem. Phys.}, 28:956--961, 1958.

\bibitem{cher1958}
M.~Cher and G.~B. Kistiakowsky.
\newblock Gaseous detonations. {XI}. double waves.
\newblock {\em J. Chem. Phys.}, 29:506--511, 1958.

\bibitem{kistiakowsky1959}
G.~B. Kistiakowsky and F.~D. Tabbutt.
\newblock Gaseous detonations. {XII}. rotational temperatures of the hydroxyl
  free radicals.
\newblock {\em J. Chem. Phys.}, 30:577--581, 1959.

\bibitem{kistiakowsky1961}
G.~B. Kistiakowsky and R.~K. Lyon.
\newblock Gaseous detonations. {XIII}. rotational and vibrational distributions
  of {OH} radicals.
\newblock {\em J. Chem. Phys.}, 35:995--998, 1961.

\bibitem{sokolik1963}
A.~S. Sokolik.
\newblock {\em Self-ignition, Flame and Detonation in Gases}.
\newblock Israel Program for Scientific Translations, Jerusalem, 1963.

\bibitem{stokes}
S.~S.~S. Cardoso, J.~H.~E. Cartwright, H.~E. Huppert, and C.~Ness.
\newblock Stokes at 200: a celebration of the remarkable achievements of {Sir
  George Gabriel Stokes} two hundred years after his birth.
\newblock {\em Phil. Trans. Roy. Soc. A}, 378:20190505, 2020.

\bibitem{voevodsky1965}
V.~V. Voevodsky and R.~I. Soloukhin.
\newblock On the mechanism and explosion limits of hydrogen-oxygen chain
  self-ignition in shock waves.
\newblock {\em Proc. Combust. Inst}, 10(1):279--283, 1965.

\bibitem{meyer1971}
J.~W. Meyer and A.~K. Oppenheim.
\newblock On the shock-induced ignition of explosive gases.
\newblock In {\em Symposium (International) on Combustion}, volume~13, pages
  1153--1164. Elsevier, 1971.

\bibitem{boeck2017}
L.~R. Boeck, R.~Mevel, and T.~Sattelmayer.
\newblock Models for shock-induced ignition evaluated by detailed chemical
  kinetics for hydrogen/air in the context of deflagration-to-detonation
  transition.
\newblock {\em Journal of Loss Prevention in the Process Industries},
  49:731--738, 2017.

\bibitem{willand1998}
J.~Willand, R.-G. Nieberding, G.~Vent, and C.~Enderle.
\newblock The knocking syndrome --- its cure and its potential.
\newblock {\em J. Fuels and Lubricants}, 107:1122--1129, 1998.

\bibitem{taveau2012}
J.~Taveau.
\newblock The {Buncefield} explosion: Were the resulting overpressures really
  unforeseeable?
\newblock {\em Process Safety Progress}, 31:55--71, 2012.

\bibitem{campbell2008}
A.~N. Campbell, S.~S.~S. Cardoso, and A.~N. Hayhurst.
\newblock Oscillatory and nonoscillatory behavior of a simple model for cool
  flames, {Sal'nikov's} reaction, {P} $\to$ {A} $\to$ {B}, occurring in a
  spherical batch reactor with varying intensities of natural convection.
\newblock {\em Combustion and Flame}, 154:122--142, 2008.

\bibitem{liu2008}
T.-Y. Liu, A.~N. Campbell, S.~S.~S. Cardoso, and A.~N. Hayhurst.
\newblock Effects of natural convection on thermal explosion in a closed
  vessel.
\newblock {\em Phys. Chem. Chem. Phys.}, 10:5521--5530, 2008.

\bibitem{cardoso2004_1}
S.~S.~S. Cardoso, P.~C. Kan, K.~K. Savjani, A.~N. Hayhurst, and J.~F.
  Griffiths.
\newblock The effect of natural convection on the gas-phase {Sal'nikov}
  reaction in a closed vessel.
\newblock {\em Phys. Chem. Chem. Phys.}, 6:1687--1696, 2004.

\bibitem{cardoso2004_2}
S.~S.~S. Cardoso, P.~C. Kan, K.~K. Savjani, A.~N. Hayhurst, and J.~F.
  Griffiths.
\newblock The computation of the velocity, concentration, and temperature
  fields during a gas-phase oscillatory reaction in a closed vessel with
  natural convection.
\newblock {\em Combustion and Flame}, 136:241--245, 2004.

\bibitem{liu2010}
T.-Y. Liu, A.~N. Campbell, A.~N. Hayhurst, and S.~S.~S. Cardoso.
\newblock On the occurrence of thermal explosion in a reacting gas: The effects
  of natural convection and consumption of reactant.
\newblock {\em Combustion and Flame}, 157:230--239, 2010.

\bibitem{azevedo2014}
F.~Gon{\c{c}}alves~de Azevedo, J.~F. Griffiths, and S.~S.~S. Cardoso.
\newblock Effects of kinetic and transport phenomena on thermal explosion and
  oscillatory behaviour in a spherical reactor with mixed convection.
\newblock {\em Phys. Chem. Chem. Phys.}, 16:23365--23378, 2014.

\bibitem{liu2013}
T.-Y. Liu and S.~S.~S. Cardoso.
\newblock Effects of combined natural and forced convection on thermal
  explosion in a spherical reactor.
\newblock {\em Combustion and Flame}, 160:191--203, 2013.

\bibitem{oran}
E.~S. Oran and A.~M. Khokhlov.
\newblock Deflagrations, hot spots, and the transition to detonation.
\newblock {\em Phil. Trans. Roy. Soc. A}, 357:3539--3551, 1999.

\bibitem{kundu2016}
S.~Kundu, J.~Zanganeh, and B.~Moghtaderi.
\newblock A review on understanding explosions from methane--air mixture.
\newblock {\em Journal of Loss Prevention in the Process Industries},
  40:507--523, 2016.

\bibitem{wang2017}
B.~Wang, Z.~Rao, Q.~Xie, P.~Wola{\'n}ski, and G.~Rarata.
\newblock Brief review on passive and active methods for explosion and
  detonation suppression in tubes and galleries.
\newblock {\em Journal of Loss Prevention in the Process Industries},
  49:280--290, 2017.

\bibitem{chamberlain2019}
G.~Chamberlain, E.~Oran, and A.~Pekalski.
\newblock Detonations in industrial vapour cloud explosions.
\newblock {\em Journal of Loss Prevention in the Process Industries},
  62:103918, 2019.

\bibitem{xu2020}
Y.~Xu, Y.~Huang, and G.~Ma.
\newblock A review on effects of different factors on gas explosions in
  underground structures.
\newblock {\em Underground Space}, 5:298--314, 2020.

\bibitem{oran2020}
E.~S. Oran, G.~Chamberlain, and A.~Pekalski.
\newblock Mechanisms and occurrence of detonations in vapor cloud explosions.
\newblock {\em Progress in Energy and Combustion Science}, 77:100804, 2020.

\bibitem{poludnenko2019}
A.~Y. Poludnenko, J.~Chambers, K.~Ahmed, V.~N. Gamezo, and B.~D. Taylor.
\newblock A unified mechanism for unconfined deflagration-to-detonation
  transition in terrestrial chemical systems and type ia supernovae.
\newblock {\em Science}, 366:eaau7365, 2019.

\bibitem{langton2010}
R.~Langton, C.~Clark, M.~Hewitt, and L.~Richards.
\newblock Aircraft fuel systems.
\newblock {\em Encyclopedia of Aerospace Engineering}, 2010.

\bibitem{nmab1980}
{\em Materials Aspects of Inert Gas Systems for Cargo Tank Atmosphere Control}.
\newblock National Academy of Sciences, Washington, DC, 1980.

\bibitem{nilsson2013}
M.~Nilsson.
\newblock Advantages and challenges with using hypoxic air venting as fire
  protection.
\newblock {\em Fire and Materials}, 38:559--575, 2013.

\bibitem{devanney2010}
J.~Devanney.
\newblock The strange history of tank inerting.
\newblock Technical report available at http://www.c4tx.org/ctx/pub/igs.pdf,
  2010.

\bibitem{atkinson2017}
G.~Atkinson, E.~Cowpe, J.~Halliday, and D.~Painter.
\newblock A review of very large vapour cloud explosions: Cloud formation and
  explosion severity.
\newblock {\em Journal of Loss Prevention in the Process Industries},
  48:367--375, 2017.

\bibitem{domingos2013}
M.~Domingos and S.~S.~S. Cardoso.
\newblock Turbulent two-phase plumes with bubble-size reduction owing to
  dissolution or chemical reaction.
\newblock {\em J. Fluid Mech.}, 716:120--136, 2013.

\end{thebibliography}

\end{document}